\documentclass[amssymb,amsmath,superscriptaddress,nofootinbib,prl,reprint]{revtex4}
\usepackage{natbib,graphicx,subfigure,subeqnarray,fancyhdr,amsmath,epstopdf,multirow,color, mathrsfs}
\begin{document}
\renewcommand{\labelitemi}{$-$}
\newcommand{\change}[1]{{\color{black} #1}}
\newcommand{\Fc}{\mathcal{F}}\newcommand{\Rc}{\mathcal{R}}\newcommand{\dd}{\mathrm{d}}
\newcommand{\ee}{\mathrm{e}}\newcommand{\ci}{\mathrm{i}}\newcommand{\ib}{\mathbf{i}}
\newcommand{\jb}{\mathbf{j}}\newcommand{\kb}{\mathbf{k}}\newcommand{\ab}{\mathbf{a}}
\newcommand{\Fb}{\mathbf{F}}\newcommand{\fb}{\mathbf{f}}\newcommand{\Gb}{\mathbf{G}}
\newcommand{\Mb}{\mathbf{M}Ä}\newcommand{\nb}{\mathbf{n}}\newcommand{\Sb}{\mathbf{S}}
\newcommand{\Sbs}{\mathbf{S^*}}\newcommand{\Rb}{\mathbf{R}}\newcommand{\Sigb}{\boldsymbol{\Sigma}}
\newcommand{\Sigbs}{\boldsymbol{\Sigma^*}}\newcommand{\alphab}{\boldsymbol\alpha}
\newcommand{\omegab}{\boldsymbol{\omega}}
\newcommand{\sigmab}{\boldsymbol{\sigma}}
\newcommand{\epsb}{\boldsymbol{\epsilon}}
\newcommand{\ub}{\mathbf{u}}
\newcommand{\eb}{\mathbf{e}}\newcommand{\vv}[1]{\underline{#1}}\newcommand{\ev}{\vv{e}}
\newcommand{\rv}{\vv{r}}\newcommand{\TT}[1]{\underline{\underline{#1}}}\newcommand{\omb}{\mathbf{\omega}}
\newcommand{\Ub}{\mathbf{U}}\newcommand{\xb}{\mathbf{x}}\newcommand{\rb}{\mathbf{r}}
\newcommand{\ssb}{\mathbf{s}}\newcommand{\Xb}{\mathbf{X}}\newcommand{\Pe}{\mbox{Pe}}
\newcommand{\mean}[1]{\langle #1\rangle}
\newcommand{\ddp}{[p]^\pm}\newcommand{\taub}{\mbox{\boldmath$\tau$}}\newcommand{\Fr}{\mbox{\textit{Fr}}}
\let\grad\nabla\newcommand{\z}{\zeta}\newcommand{\kk}{\kappa}\newcommand{\tkk}{\tilde{\kappa}}
\newcommand{\e}{\varepsilon}\newcommand{\zb}{\bar{\zeta}}\let\grad\nabla\let\bcdot\cdot
\newcommand{\half}{{\textstyle\frac{1}{2}}}
\newcommand{\textfrac}[2]{{\textstyle\frac{#1}{#2}}}
\newcommand{\LF}[1]{{#1}^{\mathrm{LF}}}\newcommand{\Lap}[1]{{#1}^{\mathrm{L}}}
\newcommand{\ds}{*\!*}\newcommand{\cond}[2]{\frac{\mathrm{D} #1}{\mathrm{D} #2}}
\newcommand{\pard}[2]{\frac{\partial #1}{\partial #2}}\newcommand{\totd}[2]{\frac{\mathrm{d}#1}{\mathrm{d}#2}}
\newcommand{\pardd}[3]{\frac{\partial^2 #1}{\partial #2 \partial #3}}
\newcommand{\Rey}{\mbox{Re}}\newcommand{\Imag}{\mbox{Im}}
\newcommand{\Fpint}{=\!\!\!\!\!\!\!\int}
\newcommand{\txi}{\tilde\xi}\newcommand{\dxi}{\delta\xi}
\newcommand{\tpsi}{\tilde\psi}\newcommand{\dpsi}{\delta\psi}
\makeatletter
\def\sgn{\mathop{\operator@font sgn}}
\makeatother
\def\v{\vspace{2cm}}

\title{Geometric tuning of self-propulsion for Janus catalytic particles}

\author{S\'ebastien Michelin}
\email{sebastien.michelin@ladhyx.polytechnique.fr}
\affiliation{LadHyX -- D\'epartement de M\'ecanique, CNRS -- Ecole
  polytechnique, 91128 Palaiseau Cedex, France}
\author{Eric Lauga}
\email{e.lauga@damtp.cam.ac.uk}
\affiliation{Department of Applied Mathematics and Theoretical Physics, University of Cambridge, CB3 0WA, Cambridge, United Kingdom}

\date{\today}
\begin{abstract}
Catalytic swimmers have attracted much attention as alternatives to biological systems for examining collective microscopic dynamics and the response to physico-chemical signals.  Yet, understanding and predicting even the most fundamental characteristics of their individual propulsion still raises important challenges.    While chemical asymmetry is widely recognized as the cornerstone of catalytic propulsion, different experimental studies have reported that  particles with identical chemical properties may  propel in opposite directions. Here, we show that, beyond its chemical properties,  the detailed shape of a catalytic swimmer   plays an essential role in determining its direction of motion,     demonstrating the compatibility of the classical  theoretical framework with experimental observations.
\end{abstract}
\maketitle

The individual swimming motion of bacteria and other microscopic swimmers could go largely unnoticed because of their small size~\citep{lauga2009}. Yet, understanding the physical mechanisms underlying their  locomotion and response to biochemical and mechanical stimuli is critical as they represent a majority of the Earth's biomass and are essential to many biological processes, ranging from our own metabolism to the carbon cycle of our planet~\citep{guasto2012,doostmohammadi2012}.  In particular, the interactions between swimming cells may lead to collective behavior characterized by length scales larger than that of the individual cells, significantly modifying for example the effective properties of cell suspensions, from enhanced mixing~\citep{leptos2009} to  non-Newtonian properties~\citep{lopez2015}. 

In order to understand and characterize  the properties of such ``active fluids'', a number of controllable physical alternatives to biological systems have recently been synthesized and tested~\citep{wang2013}.   These self-propelled artificial systems, 
which also hold the promise of helping to miniaturize future biomedical applications~\citep{nelson2010}
 and can be broadly classified into two different categories: actuated and catalytic swimmers. While   actuated swimmers rely on  external fields, typically  electric \cite{bricard2013emergence}, magnetic \cite{dreyfus2005,ghosh2009} and acoustic \cite{wang2012,nadal2014}, in order to drive the actuation,  catalytic propulsion  exploits the short-range interactions of a rigid particle with the chemical ``fuel'' content of its immediate environment in order to generate autonomous propulsion.

Catalytic swimmers take a multitude of diverse forms, as demonstrated by a wealth of recent experiments. Over the last decade, pioneering work on self-propelled bi-metallic rods~\citep{paxton2004,fournierbidoz2005} has inspired many alternatives using the catalytic decomposition of hydrogen peroxyde (e.g.~partially Pt-coated colloids~\citep{howse2007} or photo-activated hematites~\citep{palacci2013}), other redox reactions~\citep{ibele2009} or decomposition of a binary mixture close to the critical point~\cite{volpe2011}. Other systems rely on heat exchanges between the particle and its environment~\citep{jiang2010,baraban2013}.

Among all these swimmers, the details of  individual propulsion mechanisms may vary and  remain in fact often poorly  understood  due to the variety \change{and complexity of the } physico-chemical mechanisms \change{involved}~\citep{duan2015,yadav2015}. Nevertheless, all catalytic swimmers share two fundamental properties, which have recently been used to propose a generic  theoretical framework for catalytic propulsion~\citep{golestanian2007}: (i) A \emph{phoretic mobility}, $M$, which is the ability to generate locally a fluid flow over the surface of the particle  from local thermodynamic gradients or electric fields along the surface;  and (ii) a physico-chemical \emph{activity}, $A$,  which modifies the environment of the particle through catalytic reactions or heat exchanges on its boundary.  Both the mobility and the activity are local physico-chemical properties of the surface of the particle and their  sign and magnitude depend on the direction of the surface  fluxes ($A$) and on  the  interaction between the surface of the particle and its physico-chemical environment ($M$)~\citep{anderson1989}.

Like biological swimmers, catalytic particles must break symmetries in order to propel in inertia-less flows~\citep{lauga2009}. Much of the current understanding of catalytic propulsion is based on the following  rationale: An asymmetric activity guarantees a polar physico-chemical environment, which  in turn, creates a directed slip flow along the surface of the  particle   providing    hydrodynamic stresses necessary to overcome viscous resistance and induce propulsion~\citep{golestanian2007,yadav2015}. This argument establishes a direct link between the polarity of the ``Janus'' particle's activity and its swimming direction \change{(e.g.~all Au-Pt colloids would swim with the platinum end forward~\citep{paxton2004})}, and was recently confirmed theoretically \change{for particles with} homogeneous mobility~\citep{nourhani2016} -- a condition     unlikely to hold for real catalytic swimmers with two chemically-distinct halves.

Recent studies have demonstrated however that this simple picture  is in  contradiction with experiments. Direct measurements have characterized the swimming  of catalytic colloids, including Pt-coated silica and polystyrene colloids in H$_2$O$_2$ solutions~\citep{ke2010,ebbens2011,valadares2010}. These have an active site (platinum) where peroxide decomposition occurs while the rest of the particle is chemically inert. The above argument would predict a uniquely-defined swimming direction, either away or toward the active Pt-site, depending only on the  relative sign of the activity and mobility.  \change{Within that framework, measured  changes in swimming speed and direction are  interpreted as changes in the chemical environment of the particle and/or of its surface chemistry 
 resulting in modifications of its mobility (for example addition of surfactants~\citep{brown2014}). Both swimming directions, e.g.~toward and away from the Pt catalytic site, are however observed in experiments for swimmers with identical surface properties and chemical conditions, that differ only by their detailed geometry~\citep{valadares2010,ke2010}, demonstrating }that surface chemistry alone is not sufficient to predict the swimming direction. This reversal in the swimming direction \change{has also been invoked to suggest} an incompatibility between these experimental observations and the classical framework of autophoresis~\citep{brown2014}.

We resolve this disagreement here by exploring the role of particle geometry in setting the direction of phoretic swimming since  the experimental studies    above differ primarily in the distinct shapes of colloidal swimmers they consider. Spherical Pt-coated silica Janus spheres  swim away from their catalytic Pt site~\citep{ke2010}, as do Pt-PS colloids~\cite{ebbens2011}, while spherical dimers consisting of a Pt sphere attached to a silica bead swim in the opposite direction~\citep{valadares2010}.  Using the classical framework of autophoresis, we demonstrate  that knowing \change{the polarity of the chemical properties alone} is not sufficient to determine the swimming direction of a catalytic particle. In fact, 
 catalytic swimmers with identical chemical properties may swim in opposite directions because of their geometrical differences.  We examine in detail the locomotion of catalytic spheres, dimers and  spheroidal colloids and show how locomotion depends on the right combination of  geometry and chemistry. 
In particular, we establish that  the swimming direction of elongated particles (rods)  is generically  opposite to that of flat colloids (disks).

\section{Results}
\subsection{Canonical framework}
Since our interest is on the role of geometry, we focus in what follows  on a generic catalytic swimming mechanism, namely self-diffusiophoresis~\citep{golestanian2007}. 
 The phoretic slip velocity along the boundary of the particle, $\ub^S(\xb)$,  is proportional to the local gradient of a solute's concentration, $C(\xb)$, along the surface as a result of short-ranged solute-particle interactions~\citep{anderson1989}. The slip velocity is thus written as $\ub^S(\xb)=M(\xb)\grad_\parallel C$, where $M$ is the phoretic mobility, an intrinsic 
  property of the surface;
 $M>0$ results from repulsive  solute-particle interactions while attractive interactions lead to   $M<0$. The solute is assumed to diffuse passively around the particle with diffusivity $D$, i.e.~satisfies $D\nabla^2 C=0$ (for small solute molecules, solute advection by the phoretic flows is negligible). On the surface of the particle, we assume that the solute is absorbed or released at a fixed rate,   $D\nb\cdot\grad C=-A(\xb)$,   where the activity $A(\xb)$ is positive for production and negative for consumption. Note that the constant-flux formulation  is fully consistent with the classical Michaelis-Menten  framework in the limit of large solute substrate concentration. This canonical autophoretic  framework and its results are  also applicable to most other autophoretic mechanisms, at least in a linearized regime, including thermophoretic and electrophoretic propulsion, which create slip flows from gradients in temperature or electric potential~\citep{anderson1989,brenner2011,bickel2013,yariv2011b}.

In the following (unless stated otherwise) we consider geometrically- and chemically-axisymmetric Janus particles of axis $\eb_z$ with one \change{active} and one inert halves. Each half has uniform, but possibly distinct, mobility and activity: ($A_1=A$, $M_1$) for the active site ($z>0$)  and ($A_2=0$, $M_2$) for the inert site ($z<0$). By symmetry, swimming occurs along the axis, with speed $\mathbf{V}=V\eb_z$.  The problem is characterized by a velocity scale $V_\textrm{ref}=AM_\textrm{rms}/D$, where $M_\textrm{rms}=\sqrt{(M_1^2+M_2^2)/2}$. Notably, $V_\textrm{ref}$ is independent of the size of the  particle~\citep{golestanian2007}. Within this framework, the reduced velocity of the catalytic particle, $U=V/V_\textrm{ref}$, is exclusively determined by the mobility ratio $M_2/M_1$ and the shape of the active colloid. Without any loss of generality, we consider below the case in which  the solute is released from, and interacts repulsively with, the active site, i.e.~we assume $A>0$, $M_1>0$. Changing the sign of one of these quantities simply reverses the swimming velocity direction~\citep{michelin2014}.  The concentration distribution around the catalytic particle and its swimming velocity are obtained analytically for three different particle shapes: (i) a sphere, (ii) a two-sphere dimer and (iii) a generic spheroidal particle (see Methods). For more general particle shapes, boundary-element methods provide a convenient framework~\citep{montenegrojohnson2015}.

\subsection{Spheres vs. dimers}

\begin{figure}[h!]
\begin{center}
\includegraphics[height=17.5cm]{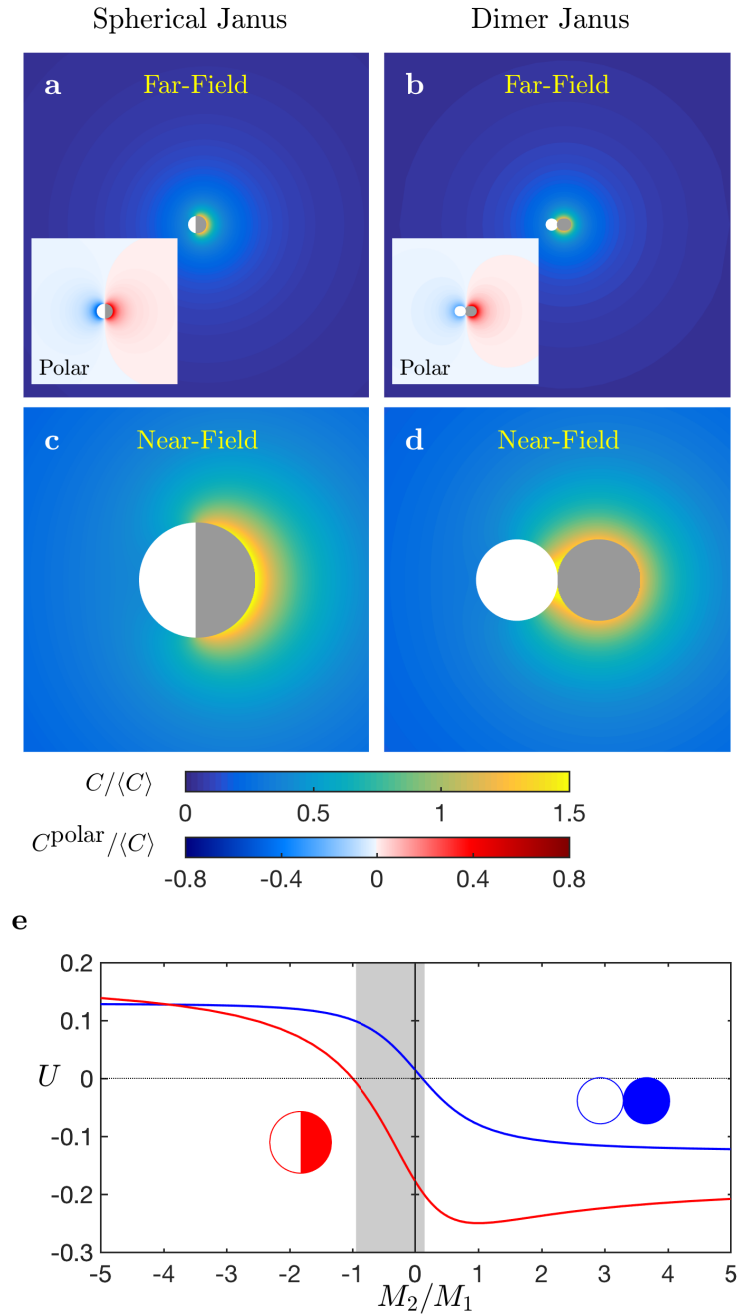}
\caption{(\textbf{a}--\textbf{b}) Far-field concentration distribution around a spherical Janus particle (\textbf{a}) and Janus dimer (\textbf{b}), each with a chemically-active right half  and a passive left half.  The concentration is scaled by its average on the particle's surface. Inset:  Anisotropic (polar) part of the concentration field, $C^\textrm{polar}$, obtained by removing the leading-order contribution of the net isotropic source. 
(\textbf{c}--\textbf{d}) Near-field  details of the concentration field. (\textbf{e})  Swimming velocity, $U$, as a function of the mobility ratio, $M_2/M_1$. For  particles similar chemically, geometry induces a velocity reversal over a finite range of mobility ratio (grey)}\label{fig:dimer}
\end{center}
\end{figure}

We first consider the spherical and dimer geometries used in experiments~\citep{valadares2010,ebbens2011}. Both particles have the same polar activity distribution
 (i.e.~right active and left inert halves). In both cases, the solute concentration can be decomposed into its isotropic and polar components as $C=C^\textrm{iso}(\xb)+C^\textrm{polar}(\xb)$. The isotropic concentration field, $C^\textrm{iso}=A_1S_1/(4\pi Dr)$, is dominant in the far-field and   corresponds to a net source of solute; it is identical for spheres and dimers of equal active surface area $S_1$. The non-isotropic part of the concentration, $C^\textrm{polar}$,  is dominated in the far-field by a source dipole whose (signed) polarity is identical for both particles since it is set by the polarity in surface chemical activity (Figure~\ref{fig:dimer} \change{a,b}).

Zooming-in on the particles reveals however a stark difference in the near-field distribution of the concentration, and in the resulting orientation of its gradient, $\grad_\parallel C$,  along the active cap  (Figure~\ref{fig:dimer} \change{c,d}).  For the spherical particle, this gradient is oriented toward the active pole where $C$ is maximum. \change{The absence of chemical activity on the passive site reduces the concentration levels in its vicinity (in comparison with a uniformly active particle that induces an isotropic concentration distribution), and this effect is minimum at the active pole.} In contrast, the gradient  takes the opposite direction \change{for the dimer geometry}, pointing toward the equatorial plane \change{where the two spheres are in contact. In that interstitial region, the diffusion of solute is predominantly two-dimensional due to confinement, which strongly limits the solute transport. The solute release being imposed by the surface activity, local gradients are enhanced to maintain the diffusive flux, resulting in increase of the local concentration near the point of contact (see Ref.~\citep{yariv2016} for a more detailed analysis of this effect).} With identical chemical properties, phoretic propulsive forces generated by the active site are therefore oriented in opposite directions, respectively away from and toward the active site for the sphere and dimer. As a result, opposite swimming velocities for these two geometries are observed  over an extended range of mobility ratio $M_2/M_1$ (Figure~\ref{fig:dimer}\change{e}). \change{These results  clearly illustrate the joint role of geometry and chemistry in setting the swimming direction of the catalytic particle. For large values of $|M_2/M_1|$, the swimming direction is entirely set by chemistry (and in particular the sphere and dimer propel in the same direction), while a geometry-induced reversal of the swimming direction is seen at intermediate values.}


\begin{figure*}[t!]
\begin{center}
\includegraphics[height=17.5cm]{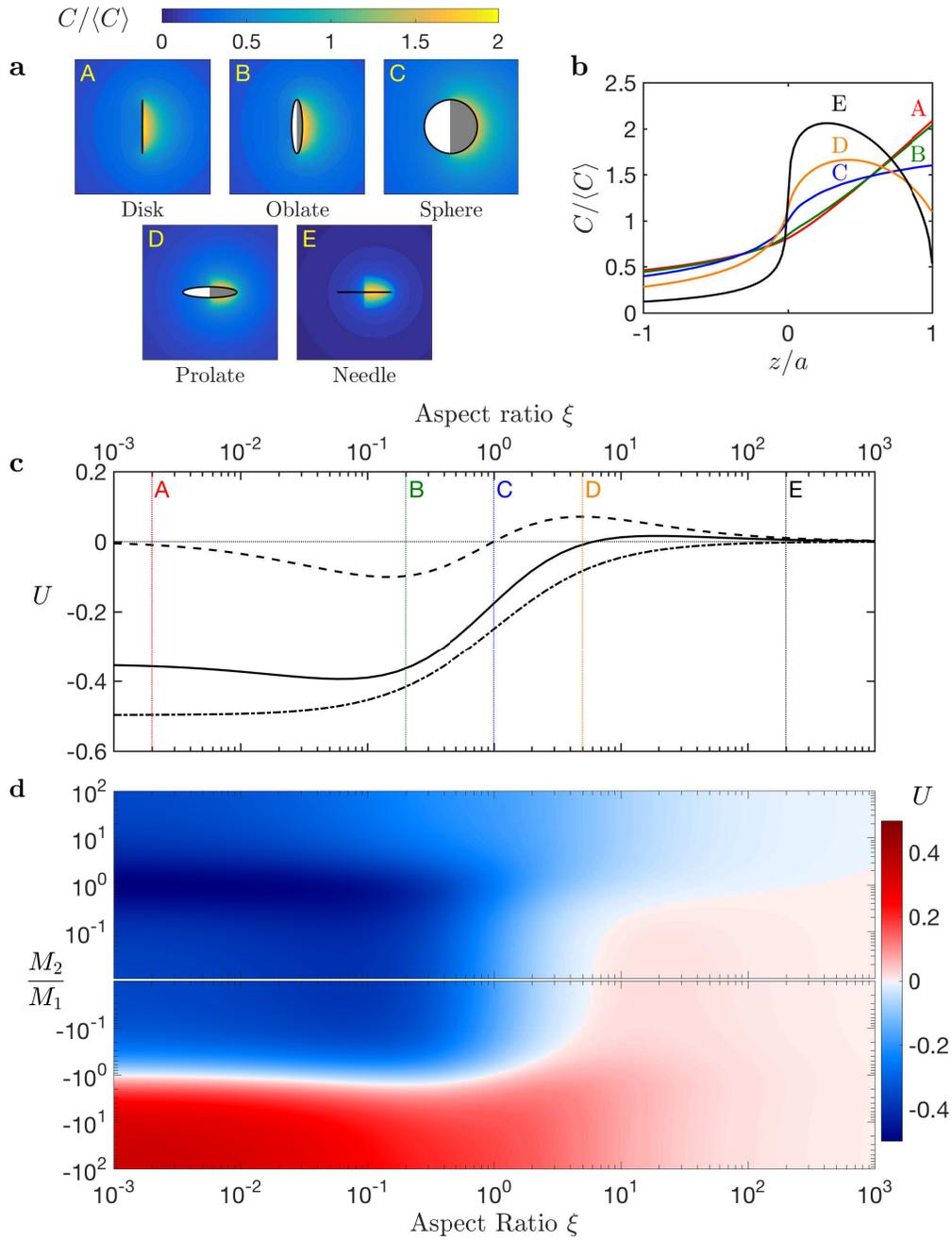}
\caption{Swimming velocity of spheroidal Janus  particles of aspect ratio $\xi$. \change{(\textbf{a}) Concentration field around representative Janus particles and (\textbf{b}) corresponding surface concentration for increasing aspect ratios $\xi$.} The right hemisphere releases a fixed flux of solute while the left hemisphere is chemically-inert. 
(\textbf{c}) Variation of the swimming velocity  with the aspect ratio $\xi$ for three different mobility distributions:  uniform mobility ($M_1=M_2$, dash-dotted line);  a fully-inert left-hand hemisphere ($M_2=0$, solid line);  opposite mobilities on both sites ($M_1=-M_2$, dashed line). The five geometries (A--E) from the top panel are indicated. 
 (\textbf{d}) Dependence of the swimming velocity $U$ with aspect ratio, $\xi$, and mobility ratio, $M_2/M_1$.}\label{fig:ell_vel_conc}
\end{center}
\end{figure*}

\subsection{The impact of the eccentricity of the particle}	
This geometric reversal of the swimming geometry stems from the detailed, local distribution of a diffusive (and therefore harmonic) field around non-spherical particles. This can be explored further by considering  spheroidal Janus particles.  This generic geometry, amenable to analytic calculations, is characterized by a single parameter, namely its aspect ratio, $\xi=a/b$, where $a$ and $b$ are the polar and equatorial radii, respectively. Varying $\xi$ allows to span a full range of axisymmetric shapes, from slender rods ($\xi\gg 1$) to flat disks ($\xi\ll 1$).

For all values of  $\xi$, the anisotropic part of the chemical signature of the particle in the far-field  is governed by the polarity of the  activity on its surface; it therefore takes a unique sign for all shapes. However, the detailed geometry controls the near-field diffusive dynamics and the surface concentration (Figure~\ref{fig:ell_vel_conc} \change{a,b}). For spheres and axisymmetric disks, the maximum concentration is reached at the active pole. \change{In contrast, for elongated and rod-like particles, the strong local curvature at the pole provides a wider solid angle available for the solute diffusion}, resulting in weaker gradients and concentration levels in the pole's vicinity. The chemical gradient along the active site is therefore reversed, \change{now pointing away from the pole and} toward  the location of the maximum concentration, which is positioned   near the equatorial plane in the limit $\xi\rightarrow\infty$ \change{(Figure~\ref{fig:ell_vel_conc} \change{a,b})}.

 As a result, the phoretic forcing of the active site points in opposite directions for rods and disks (or spheres). For solute particles releasing a repulsive solute ($A,M_1>0$), this phoretic force is always oriented toward the inert cap for oblate and spherical particles, but toward the active cap for prolate swimmers. This results in opposite swimming directions for elongated and flat particles (Figure~\ref{fig:ell_vel_conc} \change{c}) provided the mobility of the active cap dominates (in magnitude) that of the inert cap (i.e.~provided that $|M_2/M_1|<1$, Figure~\ref{fig:ell_vel_conc} \change{d}). \change{These results emphasize once again the interplay between the mobility contrast and particle shape in setting the propulsion direction.}

\subsection{Non-axisymmetric Janus particles}
\begin{figure}
\begin{center}
\includegraphics[width=.85\textwidth]{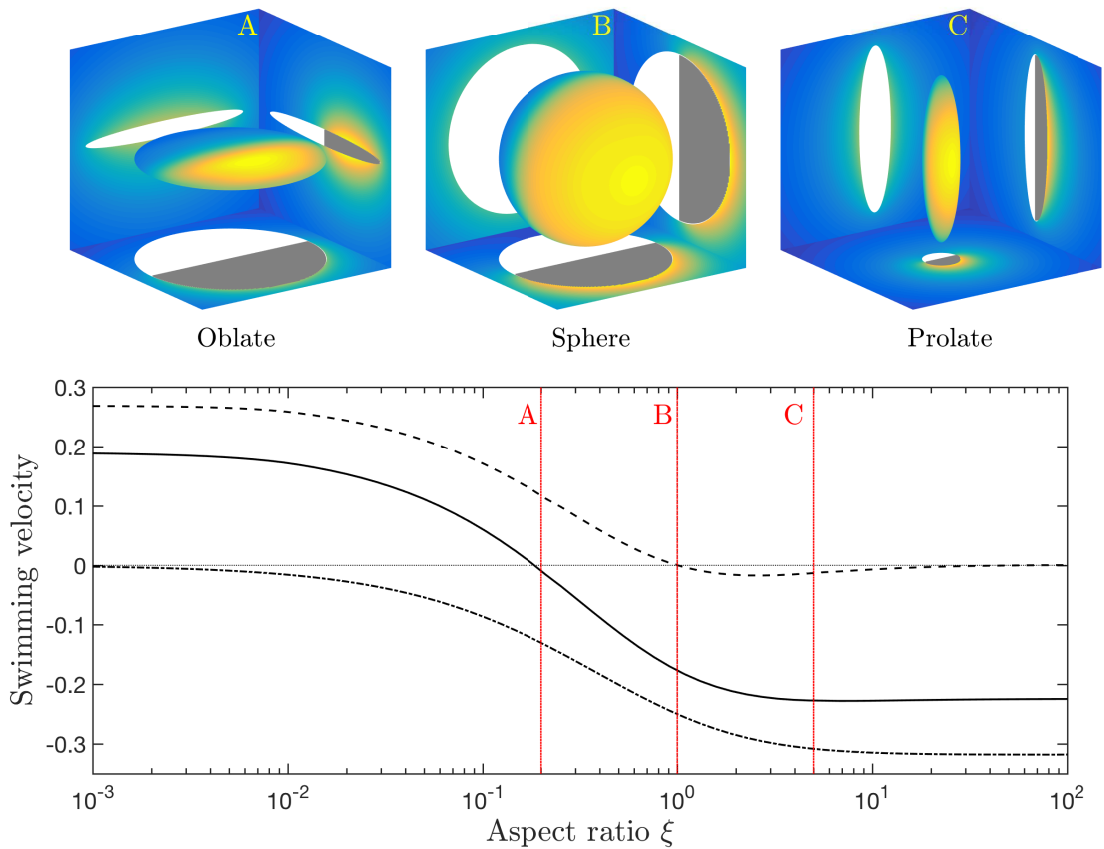}
\caption{Swimming velocity of a non-axisymmetric spheroidal Janus particle. The right half of the particle is chemically-active, while the rest of the particle is passive. Top: Solute concentration on the surface of the particle and on the planes of symmetry for three representative geometries (A--C). 
Bottom: Dependance  of the swimming velocity on the aspect ratio, $\xi$, for three different mobility distributions: uniform mobility ($M_1=M_2$, dash-dotted line); a fully-inert left-hand hemisphere ($M_2=0$, solid line);  opposite mobilities on both sites ($M_1=-M_2$, dashed line). The three geometries (A--C) from the top panels are indicated. \label{fig:camembert} }
\end{center}
\end{figure}

This critical role of the geometry of  a particle  on the sign of the swimming velocity, demonstrated above for simple axisymmetric particles, is in fact a very general feature.  In the case of more complex shape and chemical patterns, geometry  still controls, and possibly reverses, the swimming velocity.  This \change{is illustrated now} by considering chemically-asymmetric spheroidal Janus particles. These have a spheroidal geometry and are divided into two distinct chemically-homogeneous portions with a dividing plane that includes (instead of being perpendicular to) their axis of geometric symmetry (\change{Figure~\ref{fig:camembert}, top}).  Here again, rods and disks propel in opposite directions. For rods, the maximum concentration is this time reached at the center of the active site (which lies in the geometric equatorial plane), while it is shifted toward the dividing plane between active and passive sites for flattened geometries (Figure~\ref{fig:camembert}, bottom).

\begin{figure*}
\begin{center}
\includegraphics[width=.8\textwidth]{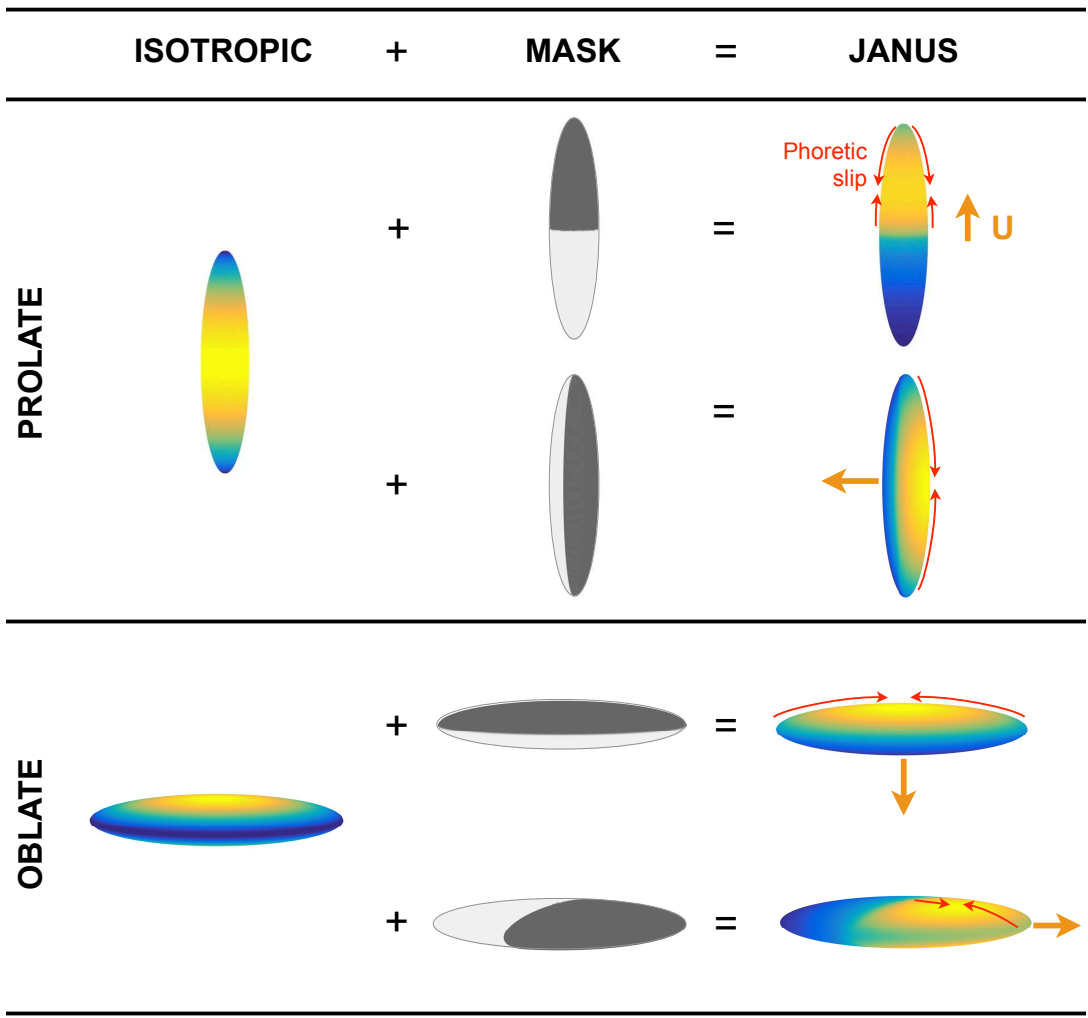}
\caption{
Effect of shape on the concentration of a diffusive species around prolate (top) and oblate (bottom) phoretic particles. 
Left: Surface concentration  for a uniform particle activity ($A=1$ everywhere on the surface) \change{with maximum  concentration obtained in regions of small curvature and minimum where the mean curvature is large}.  
\change{Center: Janus particles may be   modeled by adding Janus ``masks'' on the uniformly-active cases, to account for the effect of the passive site. 
 Right:   Surface concentration obtained for both axisymmetric and non axisymmetric Janus particles showing that  the location of the concentration maxima (and therefore the slip velocity, red arrows, and net swimming, orange arrows) is qualitatively predicted by  combining the mask with the isotropic case to model the absence of activity on one half of the particle.} The cases illustrated here are for  $A$ and $M_1$ positive and $M_2=0$.
}\label{fig:sketch_uniform}
\end{center}
\end{figure*}

We thus showed that the argument linking   the  swimming velocity of a catalytic particle with the polarity of its activity does  not hold. Is it  however possible to rationalize our new results into an intuitive prediction? In the case of a spheroidal particle, much can in fact be learnt from the distribution of a diffusive solute around a simple chemically-homogeneous active particle. These particles release (or consume) solutes \change{uniformly}  on their entire surface, and as a result cannot swim by symmetry.  For such shapes,  the maximum concentration is achieved in the regions with lowest curvature while minima are obtained in the sharpest regions \change{(this is illustrated in Figure~\ref{fig:sketch_uniform}, left). This location of the maxima and minima is  a reverse tip effect which is associated with the ability of the solute to diffuse more efficiently outward near the regions of high curvature -- a classical feature of the solutions to Laplace's equation.  We then consider a combination of these isotropic solutions with  four possible Janus ``masks'',  as illustrated in Figure~\ref{fig:sketch_uniform}. These geometric masks, which account for the absence of activity on one half of the particle, modify the concentration distributions obtained for uniformly-active particles.  Taking into account the Janus chemical patterning of the activity is equivalent to neutralizing the release of solute on one half of the particle, resulting in a reduction of the concentration field in that region, and in its immediate vicinity. This simple approach  allows thus to predict  qualitatively 
 the correct location of the concentration extrema (Figure~\ref{fig:sketch_uniform}, right), the local phoretic slip velocity (thin red arrows) and therefore the locomotion velocity (thick orange arrow).}

\section{Discussion}
Beyond governing the swimming velocity of  catalytic particles, this geometric modulation of the surface concentration field,  and thus of the phoretic slip distribution, impacts the details of the hydrodynamic flows  generated by the particles and how these affect the dynamics of other particles.  One of these characteristics is effective particle stresslets, which  quantify  their slowest-decaying hydrodynamic signature and their contribution to the bulk stress in a suspension.  For a fixed surface chemistry, the particle geometry can modify and  reverse the hydrodynamic signature from a puller-type swimmer (i.e.~an active particle that swims by pulling itself forward, similarly to algae such as \emph{Chlamydomonas Reinhardii}) to a pusher-type swimmer (i.e.~one that swims by pushing the fluid in its wake, similarly to flagellated  bacteria such as \emph{Escherichia coli})~\citep{lauga2016b}.  In designing Janus catalytic particles, geometric modulation  provides therefore an alternative design route  to chemical modulation 
 for synthesizing artificial microswimmers with tailored hydrodynamic properties.

\section{Methods}
\subsection{Diffusion and locomotion}
Finding the swimming velocity of the catalytic particle first requires solving in the fluid domain, $\Omega_f$, for the concentration of the diffusing solute relative to its far-field constant value, $C(\xb)=c-c_\infty$, subject to the activity-induced flux condition at the particle's surface, $\partial \Omega_f$:
\begin{equation}\label{eq:laplace}
\nabla^2 C=0\quad \textrm{in   }\Omega_f,\qquad D\nb\cdot\nabla C=-A(\xb)\quad\textrm{on }\partial\Omega_f.
\end{equation}

The phoretic slip on the boundary of the particle is then given by $\ub^S=M(\xb)\nabla_\parallel C$ 
 and using the reciprocal theorem for Stokes flows, the swimming velocity of the force- and torque-free particle along an axis $\mathbf{d}$ is obtained as~\citep{stone1996}
\begin{equation}
\Ub\cdot\mathbf{d}=\int_{\partial\Omega_f}\ub^S\cdot\boldsymbol{\tilde\sigma}\cdot\nb\dd S,\label{eq:lrt}
\end{equation}
with $\boldsymbol{\tilde\sigma}$ the stress tensor associated with the flow around the torque-free rigid particle of identical geometry pulled by a unit force along $\mathbf{d}$. Note that for axisymmetric swimmers, $\mathbf{d}$ will be chosen along the axis of symmetry $\mathbf{d}=\eb_z$.

\subsection{Spherical particle}
The axisymmetric concentration field can be decomposed into spherical harmonics and the Laplace problem, Eq.~\eqref{eq:laplace}, is solved as~\citep{golestanian2007}
\begin{equation}
c(r,\mu)=\frac{a}{D}\sum_{n=0}^\infty\left(\frac{a}{r}\right)^{n+1}\frac{(2n+1)P_n(\mu)}{2(n+1)}\int_{-1}^1 A(\zeta)P_n(\zeta)\dd\zeta,\label{eq:c_sphere}
\end{equation}
with $P_n(x)$ the Legendre polynomials of degree $n$ and $\mu\equiv \cos\theta$ in polar spherical coordinates. For a spherical particle translating along $\mathbf{d}=\eb_z$, $\boldsymbol{\tilde\sigma}\cdot\nb=-\eb_z/(4\pi a^2)$. For hemispheric Janus particles, substitution into Eqs.~\eqref{eq:c_sphere} and \eqref{eq:lrt} provides the final result~\citep{golestanian2007}
\begin{equation}
U=\frac{(A_2-A_1)(M_1+M_2)}{8D}\cdot
\end{equation}

\subsection{Spherical dimer}
A spherical dimer consists of two spheres of radius $R$ (one active and one passive) separated by a distance $d\ll R$. This geometry and the solution framework was already investigated using bi-spherical polar coordinates~\citep{popescu2011,michelin2015a,reigh2015}. The cylindrical coordinates $(\rho,\theta,z)$ are mapped onto the bi-spherical coordinates $(\tau,\mu,\theta)$ by 
\begin{equation}
(\rho,z)=\frac{a}{\cosh\tau-\mu}(\sqrt{1-\mu^2},\sinh\tau).
\end{equation}
The surface of the two spheres is $\tau=\pm\tau_0$, and $(a,\tau_0)$ are defined from $(R,d)$ as $a=R\sinh\tau_0$ and $d=2R(\cosh\tau_0-1)$. The general solution of Laplace's equation vanishing at infinity is
\begin{equation}
C(\tau,\mu)=\sqrt{\cosh\tau-\mu}\sum_{n=0}^\infty c_n(\tau)P_n(\mu),
\end{equation} 
with 
\begin{equation}
c_n(\tau)=a_n\ee^{\left(n+\frac{1}{2}\right)(\tau-\tau_0)}+b_n\ee^{-\left(n+\frac{1}{2}\right)(\tau+\tau_0)}.
\end{equation}
The constants $a_n$ and $b_n$ are found by projecting the flux boundary conditions on the surface of the active and passive spheres~\citep{michelin2015a}
\begin{equation}
\pard{c}{\tau}(\tau_0,\mu)=\frac{aA}{D(\cosh\tau_0-\mu)},\quad \pard{c}{\tau}(-\tau_0,\mu)=0.
\end{equation}
The dual flow field required for the reciprocal theorem formulation is obtained in a similar way by noting that the general axisymmetric solution of Stokes' equations with the proper decay at infinity can be found in bi-spherical coordinates in terms of a streamfunction $\tilde\psi$,
\begin{equation}
\tilde\ub=(\cosh\tau-\mu)^2\left[\pard{\tilde\psi}{\mu}\eb_\tau-\frac{1}{\sqrt{1-\mu^2}}\pard{\tilde\psi}{\tau}\eb_\mu\right],
\end{equation}
and 
\begin{eqnarray}
\tilde\psi(\tau,\mu)&=&\tilde{U}(\cosh\tau-\mu)^{-3/2}\sum_{n=1}^\infty(1-\mu^2)P_n'(\mu)U_n(\tau),\\
U_n(\tau)&=&\alpha_n\ee^{\left(n+\frac{3}{2}\right)(\tau-\tau_0)}+\beta_n\ee^{-\left(n+\frac{3}{2}\right)(\tau+\tau_0)}\nonumber\\
&&+\gamma_n\ee^{\left(n-\frac{1}{2}\right)(\tau-\tau_0)}+\delta_n\ee^{-\left(n-\frac{1}{2}\right)(\tau+\tau_0)},
\end{eqnarray}
and $\tilde{U}$ is a constant such that the total hydrodynamic drag on the dimer is $-\eb_z$. The set of constants $(\alpha_n,\beta_n,\gamma_n,\delta_n)$ is found numerically by applying the no-slip boundary condition $\tilde\ub=\tilde{U}\eb_z$ on the surface of each sphere $\tau=\pm\tau_0$. The stress force of the dual rigid problem $\boldsymbol{\tilde\sigma}\cdot\nb$ can then be found explicitly~\citep{michelin2015a}.

\subsection{Spheroids}
Focusing now on spheroidal particles of aspect ratio $\xi=a/b$ with $a$ and $b$ the polar and equatorial radii respectively, the problem formulated in Eqs.~\eqref{eq:laplace}--\eqref{eq:lrt} can be solved in spheroidal polar coordinates $(\tau,\zeta,\theta)$ defined by $(\rho,z)=k(\sqrt{\tau^2\mp 1}\sqrt{1-\zeta^2},\tau\zeta)$, with $\mp$ referring to prolate and oblate spheroids, respectively. The surface of the spheroid corresponds to $\tau=\tau_0=\xi/\sqrt{|\xi^2-1|}$ and the scaling factor $k$ is defined as $k=a/\tau_0$. The general solution of Laplace's equation vanishing at infinity is
\begin{equation}
C(\tau,\zeta,\theta)=\sum_{n=0}^\infty\sum_{m=0}^{n}P_n^m(\zeta)C_n^m(\tau)(c_n^m\cos m\theta+d_n^m\sin m\theta),
\end{equation}
with $P_n^m$ the associated Legendre polynomials, and $C_n^m(\tau)=Q_n^m(\tau)$ or $C_n^m(\tau)=Q_n^m(\ci\tau)$ for prolate and oblate spheroids, respectively, with $Q_n^m$ the associated Legendre function of the second kind~\citep{abramowitz1964}. The coefficients $c_n^m$ and $d_n^m$ are obtained by projecting the flux boundary condition on the particle's surface $\partial C/\partial \tau (\tau_0)=-kA(\zeta,\theta)\sqrt{\zeta^2+\xi^2(1-\zeta^2)}/D$~\citep{lauga2016b}
\begin{align}
\left(\begin{array}{c} c_n^m\\d_n^m\end{array}\right)=&-\frac{\gamma_m k(2n+1)(n-m)!}{2\pi(n+m)!C_n^{m'}(\tau_0)}\times\nonumber\\
\int_{-1}^1\int_0^{2\pi}&A(\zeta,\theta)\left(\begin{array}{c}\cos m\theta\\ \sin m\theta\end{array}\right)\sqrt{\zeta^2+\xi^2(1-\zeta^2)}P_n^m(\zeta)\dd\theta\dd\zeta,\label{eq:conc_spheroid}
\end{align}
with $\gamma_0=1/2$ and $\gamma_m=1$ otherwise.
This solution obviously simplifies significantly in the case of an axisymmetric surface chemistry with $c_n^m=d_n^m=0$ for $m\neq 0$.

The dual rigid body problem can be solved explicitly using the classical solution of Oberbeck for a spheroid translating along $\mathbf{d}$~\citep{lamb1932,happel1965}, and the tangential part of the dual stress field is obtained as
\begin{equation}
\boldsymbol{\tau}\cdot\boldsymbol{\tilde\sigma}\cdot\nb=-\frac{|\xi^2-1|\boldsymbol\tau\cdot\mathbf{d}}{4\pi k^2\sqrt{\zeta^2+\xi^2(1-\zeta^2)}},
\end{equation}
where $\boldsymbol\tau=\mathbf{I}-\nb\nb$ is the projection operator on the surface $\partial\Omega_f$. The reciprocal theorem Eq.~\eqref{eq:lrt} then takes the simple form
\begin{equation}
\Ub=-\frac{1}{4\pi}\int_{-1}^1\int_0^{2\pi}\ub^S\dd\zeta\dd\theta.
\end{equation}
The translation velocity of the spheroid along and orthogonally to its axis of symmetry are then obtained from $C$ as
\begin{align}
U_z=\frac{\sqrt{|\xi^2-1|}}{4\pi k}&\int_{-1}^1\int_0^{2\pi}M(\zeta,\theta)\pard{C}{\zeta}\frac{\xi(1-\zeta^2)}{\zeta^2+\xi^2(1-\zeta^2)}\dd\theta\dd\zeta,\label{eq:uz_spheroid}\\
\Ub_\perp=\frac{\sqrt{|\xi^2-1|}}{4\pi k}&\int_{-1}^1\int_0^{2\pi}\left[\frac{\zeta\sqrt{1-\zeta^2}}{\zeta^2+\xi^2(1-\zeta^2)}\pard{C}{\zeta}\eb_\theta\times\eb_z+\frac{1}{\sqrt{1-\zeta^2}}\pard{C}{\theta}\eb_\theta\right]\dd\zeta\dd\theta.\label{eq:ux_spheroid}
\end{align}
Substitution of $A(\zeta,\theta)$ and $M(\zeta,\theta)$ into Eq.~\eqref{eq:conc_spheroid}, \eqref{eq:uz_spheroid} and \eqref{eq:ux_spheroid} provides an explicit expression for $\mathbf{U}$.

\section*{Funding}
This project has received funding from the European Research Council (ERC) under the European
Union's Horizon 2020 research and innovation programme under grant agreements 714027 (S.M) and 682754 (E.L.)

\section*{Author contributions}
S.M. and E.L. designed research. S.M. performed the numerical simulations. S.M and E.L. analysed data and wrote the paper.

\section*{Competing Financial Interests}
The authors have no competing interests.


\end{document}